# GENERALIZED DC LOOP CURRENT ATTACK AGAINST THE KLJN SECURE KEY EXCHANGE SCHEME

**Mutaz Y. Melhem, Laszlo B. Kish**

*Texas A&M University, Department of Electrical and Computer Engineering, College Station, TX 77843-3128, USA*
*(✉ yar111@tamu.edu, Laszlokish@tamu.edu, +1 979 847 9071)*

**Abstract**

A new attack against the Kirchhoff-Law-Johnson-Noise (KLJN) secure key distribution system is studied with unknown parasitic DC-voltage sources at both Alice's and Bob's ends. This paper is the generalization of our earlier investigation with a single-end parasitic source. Under the assumption that Eve does not know the values of the parasitic sources, a new attack, utilizing the current generated by the parasitic dc-voltage sources, is introduced. The attack is mathematically analyzed and demonstrated by computer simulations. Simple defense methods against the attack are shown. The earlier defense method based solely on the comparison of current/voltage data at Alice's and Bob's terminals is useless here since the wire currents and voltages are equal at both ends. However, the more expensive version of the earlier defense method, which is based on in-situ system simulation and comparison with measurements, works efficiently.

## 1. Introduction

The Kirchhoff-Law-Johnson-Noise (KLJN) secure key distribution scheme [1-26], see Figure 1, is based on the properties of thermal noise of the resistors. Once it was introduced in [2], it had the promise to provide robust, fast and unconditionally secure communications at low cost, and all the advantages of chip integrability. The information channel is composed of a wire. In the core of the simplest scheme, each communicator contains a switch that, at the beginning of each bit exchange cycle, randomly alternates between two resistors $R_\text{L}$ & $R_\text{H}$, where we suppose $R_\text{H} > R_\text{L}$, see Figure 1. Since, in the system, there are four different combinations of $R_\text{A}$ & $R_\text{B}$, that is, $R_\text{L}R_\text{L}$ (LL), $R_\text{H}R_\text{H}$ (HH), $R_\text{L}R_\text{H}$ (LH) and $R_\text{H}R_\text{L}$ (HL), the KLJN scheme randomly alternates between four different situations. The first and second letters in the parentheses refer to the connected resistors at Alice's and Bob's ends, respectively. Note the practical scheme consists of many more elements such as noise generators, line filters, and measurement and compensation systems to secure the privacy for arbitrary situations. The power spectral density of the voltage in the wire can be written as:

$$S_{u,w}(f) = 4kT \frac{R_\text{A} R_\text{B}}{R_\text{A} + R_\text{B}} \qquad (1)$$

While the power spectral density of the current in the wire can be written as:

$$S_{i,w\|}(f) = \frac{4kT}{R_\text{A} + R_\text{B}} \qquad (2)$$

Here $T$ is the effective noise temperature temperature (which can be much larger than room temperature when external voltage generators are used), $k$ is the Boltzmann's constant, $R_A$ and $R_B$ are the connected resistances at Alice's and Bob's ends, respectively, where $R_A, R_B \in \{R_L, R_H\}$. On the other hand, each switching cycle, Alice and Bob will measure the wire's $S_{u,w}(f)$ and $S_{i,w}(f)$, and by using either (1) or (2), each party will compensate it's resistance value and the measured $S_{u,w}(f)$ or $S_{i,w}(f)$ in the left hand side of the used equation, so each equation will end with only one variable, which is the resistance of the other party. Then they will solve the resultant equation to get the other party resistant. To guarantee total security of the key, both parties will ignore any cycle where $R_A \& R_B$ are equal because Eve can know the values of the operating resistances by solving (1) and (2), while the LH and HL situations will be considered for the Key Exchange, since in these situations Eve cannot infer which resistance is operating in either side. A general proof about the unconditional security of the Kirchhoff-Law-Johnson-Noise (KLJN) was introduced in [2].

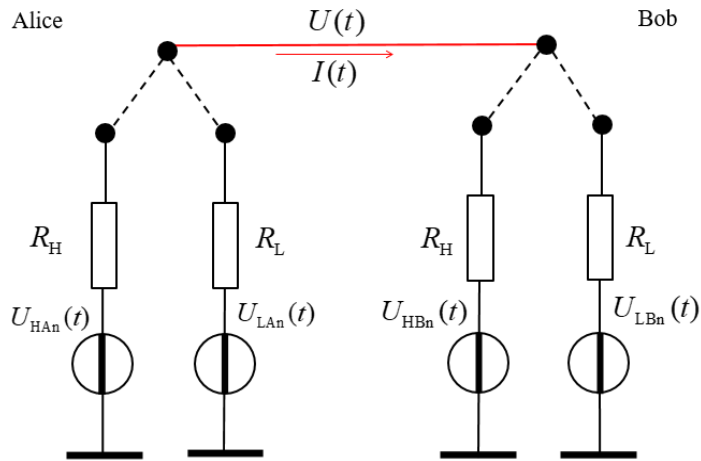

**Fig. 1.** The core of the KLJN system. $U_{HAn}(t)$, $U_{LAn}(t)$, $U_{HBn}(t)$ and $U_{LBn}(t)$ are either the natural (thermal) Johnsons noises sources of the related resistances or artificial noise generators with high voltages. Alice and Bob measures the voltage and the current $U(t)$ and $I(t)$ in the wire, and use them in evaluating the power density spectra $S_{u,w}(f)$ and $S_{i,w}(f)$, respectively.

Several attacks against the KLJN scheme were proposed against the KLJN scheme. None of them could impair the unconditional security of the scheme, since either there was a proper defense scheme against or attack was based on misconceptions [3-20, 27-30], including experimental errors. On the other hand, passive attacks against practical KLJN systems deviating from the ideal assumptions of the scheme [2] were discussed in several papers [4-13]. For example, parasitic or non-ideal features such as transients, wire's resistance, cable capacitance, temperature differences, delay effects and transients, etc. are such practical issues. In each case, these discussions ultimately confirmed the security of the practical KLJN with proper parameter choices, defense circuitries, compensation techniques, or privacy amplifications [1]..

Recently, a new attack scheme assuming the existence of single parasitic DC source at either communicating party was introduced [14]. DC current based ground loops or similar non-idealities can yield such situation. In [14] to demonstrate the security vulnerability, the following assumptions were made:

i) The parasitic DC source was located at only one of the end of the wire channel;

ii) Eve knew the location of this parasitic DC source.

iii) Eve knew the value of the voltage of this DC source.

The above assumptions [14] were in line with Kerckhoffs's principle of security [1], that is, all the essential details of the secure communication system will eventually be known by the adversary except the momentary key [1].

In the present paper, we make Eve's job much more complicated by adding an arbitrary second generator assuming that there was not yet enough time for Eve to utilize Kerckhoffs's principle of security [1], see below.

**2. The generalized DC ground loop situation**

In [14], we assumed a ground loop situation with a single, known DC voltage generator located at Alice's end and demonstrated the resulting information leak. In the present paper, the generalized and most common practical situation is studied with two unknown DC voltage generators of arbitrary polarity that are located at Alice's and Bob's sides, see Figure 2.

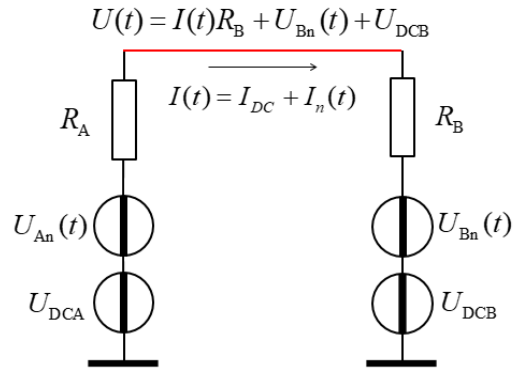

**Fig. 2.** The KLJN system in the generalized ground loop situation $U_{Bn}(t)$ & $U_{An}(t)$ are the thermal voltage noise sources associated with $R_A$ & $R_B$, respectively, where $R_A$ & $R_B \in \{R_L; R_H\}$. $U_{DCA}$ & $U_{DCB}$ are the ground loop DC voltage sources, and $U(t)$ & $I(t)$ are the wire voltage and wire current, respectively.

As a preparation, we introduce the mathematical notations that are similar but more complex than in [14]. Both the voltage $U(t)$ and current $I(t)$ in the wire have a DC and an AC (stochastic, that is, noise) components, see Figure 2. The direction of the current $I(t)$ is assumed to point from Alice to Bob. Then the current in the wire can be expressed as:

$$I(t) = I_{DC} + I_n(t), \qquad (3)$$

where the DC and AC components are:

$$I_{DC} = \frac{U_{DCA} - U_{DCB}}{R_A + R_B} \quad (4)$$

and

$$I_n(t) = \frac{U_{An}(t) - U_{Bn}(t)}{R_A + R_B} \quad . \quad (5)$$

Here $U_{An}$ and $U_{Bn}$ are the effective (rms) values of the Johnson noise voltage sources with $R_A$ and $R_B$ (either physical thermal noise or external generators representing enhanced effective temperature [1-4]) with $U_{An} \in \{U_{LAn}; U_{HAn}\}$ and $U_{Bn} \in \{U_{LBn}; U_{HBn}\}$. The voltage on the wire can be written as:

$$U(t) = I(t)R_B + U_{Bn}(t) + U_{DCB} \quad . \quad (6)$$

From (3) and (6) we obtain:

$$U(t) \equiv U_{DCw} + U_{ACw}(t) = I_{DC}R_B + U_{DCB} + I_n(t)R_B + U_{Bn}(t) \quad (7)$$

Where $U_{DCw}$ and $U_{ACw}(t)$ are the DC and AC voltage components on the wire, see Figure 2:

$$U_{DCw} = I_{DC}R_B + U_{DCB} = \frac{R_B U_{DCA} + R_A U_{DCB}}{R_A + R_B} \quad (8)$$

From Equation 8, it is obvious that a non-zero information leak occurs since the DC components are different in the LH and the HL bit case. Specifically, at the LH situation, that is, when $R_A = R_L$ and $R_B = R_H$, the DC component of the wire's voltage is:

$$U_{DCw} \equiv U_{LH} = \frac{U_{DCA}R_H + U_{DCB}R_L}{R_H + R_L} \quad (9)$$

while in the HL bit situation it is:

$$U_{DCw} \equiv U_{HL} = \frac{U_{DCA}R_L + U_{DCB}R_H}{R_H + R_L} \quad (10)$$

For later usage, we evaluate the average of the above-defined $U_{LH}$ and $U_{HL}$, and call this quantity *threshold voltage*, $U_{th}$:

$$U_{th} \circ \frac{U_{LH} + U_{HL}}{2} = \frac{U_{DCA} + U_{DCB}}{2} \qquad (11)$$

Moreover, we compare Equations (9) and (10), to obtain the following inequality:

$$U_{LH} {}^3 U_{HL} \text{ if } U_{DCA} {}^3 U_{DCB} \qquad (12)$$

The noise component $U_{ACw}(t)$ of $U(t)$, see Figure 2, can be written as:

$$U_{ACw}(t) = I_n(t)R_B + U_{Bn}(t) \qquad (13)$$

From (5) and (13):

$$U_{ACw}(t) \circ \frac{U_{An}(t) - U_{Bn}(t)}{R_A + R_B} R_B + U_{Bn}(t) = \frac{U_{An}(t)R_B + U_{Bn}(t)R_A}{R_A + R_B} \qquad (14)$$

Obviously, $U_{ACw}(t)$ has normal distribution, since it is the linear combination of Gaussian noises and DC values, and their power spectral density is the same in both the LH and HL cases [14]. Figures 3 and 4 illustrate the different situations of the wire voltage $U(t)$ versus the threshold voltage $U_{th}$, when $U_{DCA} > U_{DCB}$ and $U_{DCB} > U_{DCA}$. This behavior of the wire voltage is exploited in our new attack scheme to distinguish the LH and HL bit arrangements, as it will be discussed in Sections 3 and 4.

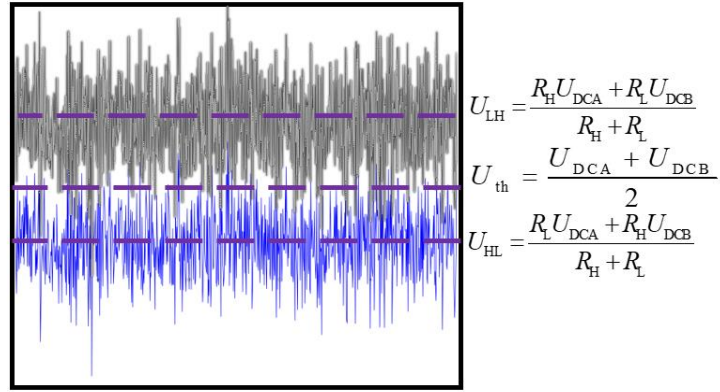

**Fig. 3**. Eves' threshold scheme to guess the bit situation LH *vs.* HL when $U_{DCA} > U_{DCB}$, and the $U_{DCA}$ and $U_{DCB}$ parasitic DC voltages are assumed positive for the purpose of illustration. $U_{LH}$ and $U_{HL}$ are the wire's DC voltages in the LH and HL situations, respectively, and $U_{th}$ is their mean.

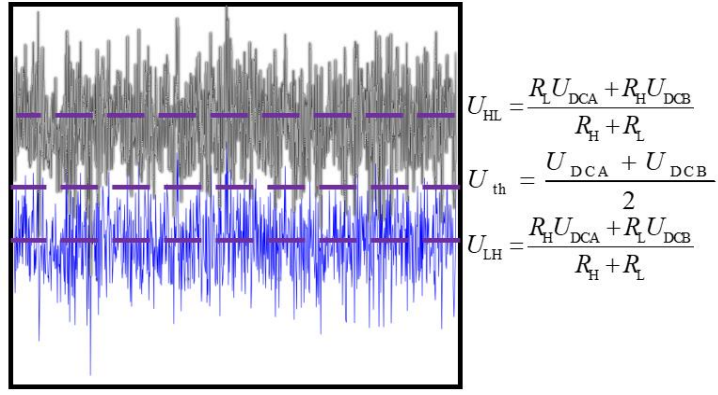

**Fig. 4**. Eves' threshold scheme to guess the bit situation LH *vs.* HL when $U_{DCB} > U_{DCA}$ and they are assumed positive for the purpose of illustration.

## 3. Eve estimates the values of the ground-loop voltages

The first step in our attack scheme is to compare the values of $U_{DCA}$ and $U_{DCB}$. Our assumption is that Eve originally does not know the values of the parasitic voltage sources, so first we introduce a technique to measure $U_{DCA}$ and $U_{DCB}$, respectively. From Equation (6), we can express $U_{DCB}$ as:

$$U_{DCB} = U(t) - I(t)R_B - U_{Bn}(t) \tag{15}$$

The above equation is useful to Eve in the LL and HH situations, where she knows the connected resistors, including $R_B$, see [14]. $U(t)$ and $I(t)$ are measurable, and even though she does not know the instantaneous signal of $U_{Bn}(t)$, she can use time averaging to produce:

$$<U_{DCB}>_t = <U(t)>_t - R_B<I(t)>_t - <U_{Bn}(t)>_t \tag{16}$$

Where $\langle \, \rangle_t$ represents the time average over a time period $t$. If $t$ is long enough, the AC components will converge to zero. From (16) we get:

$$U_{DCB} = <U(t)>_t - R_B<I(t)>_t \tag{17}$$

For finite $t$ time averages, the estimation of $U_{DCB}$ has error because the convergence to zero is incomplete. We will call this error $e_B$. Following the same procedure for $U_{DCA}$:

$$U_{DCA} = <U(t)>_t + R_A<I(t)>_t \tag{18}$$

We call the error of this estimation $e_A$. Knowing the resistance values in the LL and HH situations, Eve can estimate the values of $U_{DCB}$ and $U_{DCA}$ from (17) and (18), respectively. Equation (13) and condition $R_H > R_L$ imply that the noise voltage on the wire, and $e_A$ and $e_B$ are higher in the HH situation than in the LL situation. Accordingly, Eve should use the LL situation to estimate $U_{DCA}$ and $U_{DCB}$.

## 4. On the attack

*4.1 The attack scheme*

After Eve estimates $U_{DCA}$ and $U_{DCB}$, she conducts four more steps:

i) *Comparison of the DC voltages*: Eve uses the extracted $U_{DCA}$ and $U_{DCB}$ values to determine whether the DC voltage component in the wire is higher during the LH or the HL bit situation [see (12)]. Then Eve designs the guessing protocol discussed below.

ii) *Measurement*: During the bit exchange period (BEP), $N$ independent samples of the wire voltage are recorded by Eve.

iii) *Evaluation*: Similarly to the procedure in [14], Eve calculates the ratio $g = N^+ / N$ where $N^+$ is the number of points above $U_{th}$ and $N$ is the total number of samples.

iv) *Guessing* [based on (9-14)]: For $U_{DCA} > U_{DCB}$, Eve's guess is LH if $0.5 < g$. Conversely, her guess is HL when $g < 0.5$. For $U_{DCA} < U_{DCB}$ and $g < 0.5$, her guess is LH, and it is HL when $0.5 < g$. Regardless of the values of $U_{DCA}$ and $U_{DCB}$, Eve's decision is undetermined when $g = 0.5$.

v) *Eve's probability p of correct guessing of a bit* is the ratio of the number of correctly guessed bits $n_{cor}$ to the total number of guessed bits $n_{tot}$, $p = n_{cor} / n_{tot}$ [14]. The $p = 0.5$ situation indicates the perfect security limit [27].

*4.2 Can Eve use the DC current, instead of the DC voltage, in her attack scheme?*

To comply with the mathematical notations used in Section 2, $I_{HL}$ and $I_{LH}$ denote the DC current in the wire; and $I_{LHn}(t)$ and $I_{HLn}(t)$ the noise (AC) components, at the HL and LH bit situations, respectively. Following the voltage-based scheme, the threshold current $I_{th}$ is the average between $I_{HL}$ and $I_{LH}$. Due to Kirchhoff's loop law, both $I_{HL}$ and $I_{LH}$ are equal; hence, $I_{LH} = I_{HL} = I_{th}$. Also, $I_{LHn}(t)$ and $I_{HLn}(t)$ have the same rms values. Therefore there is no difference in the measured values that Eve could utilize for an attack.

*4.3 Impact of the difference between $U_{DCA}$ and $U_{DCB}$ on the attack's success*

Here we show that the efficiency of the attack depends on the difference between the parasitic DC voltages and not on their specific values. If $U_{DCA}$ and $U_{DCB}$ are both shifted by the same value $d$ then $U_{LH}$, $U_{HL}$, and $U_{th}$ are also shifted by $d$; see the illustration in Figs. 5 and 6.

Thus, only the difference $U_{DCA} - U_{DCB}$ of these voltages determines the efficiency of the attack, not their actual values.

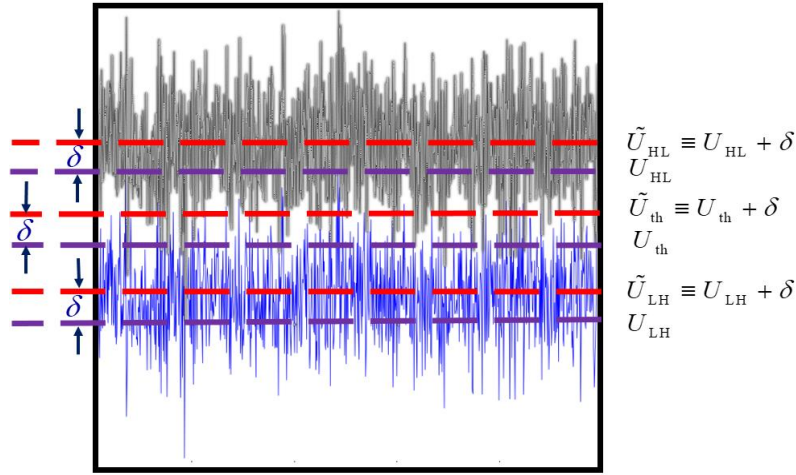

**Fig. 5**. Illustration of the DC voltage components in the LH and HL situations before and after a $\delta$ shift in the parasitic voltages, where $\tilde{U}_{DCA} = U_{DCA} + \delta$; $\tilde{U}_{DCB} = U_{DCB} + \delta$; and $\tilde{U}_{LH}$ and $\tilde{U}_{HL}$ are the resulting DC voltages in the LH and HL situations, respectively. $\tilde{U}_{th}$ is the average of $\tilde{U}_{LH}$ and $\tilde{U}_{HL}$.

In the subsequent section, the attack method is demonstrated by computer simulations.

## 5. Demonstration

Eve's correct bit guessing probability $p$ was evaluated analytically and tested by computer simulations, see Figure 6.

For the analytic evaluation we used the error function:

$$p\{U(t) \geq U_{th}\} = 0.5\left[1 - erf\left(\frac{U_{th} - U_{DCw}}{U_{eff}\sqrt{2}}\right)\right], \quad (19)$$

where $p\{U(t) \geq U_{th}\}$ is the probability of $U(t)$ to exceed $U_{th}$, and the error function is given as:

$$erf(x) = \frac{1}{\sqrt{\pi}} \int_{-x}^{x} \exp(-y^2)\, dy \quad. \quad (20)$$

And $U_{eff}$ is the effective (rms) value of the noise voltage $U_{ACw}(t)$ on the wire.

In accordance to the analysis described in Section 4.3, the results were always identical when the difference $U_{DCA} - U_{DCB}$ was the fixed, regardless of the values of $U_{DCA}$ and $U_{DCB}$. This fact confirms our theoretical result Section 4.3 that the success of the attack depends on the difference of parasitic DC sources only, and not on their actual values.

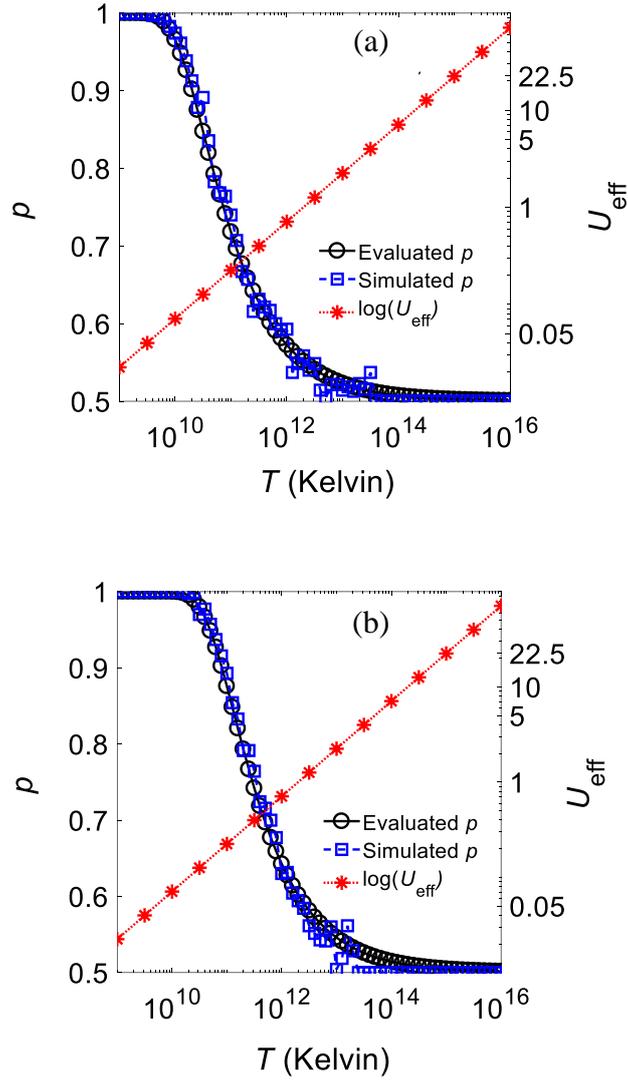

**Fig. 6**. Eve's correct bit guessing probability ($p$) versus temperatures at $U_{DCA} - U_{DCB}$ differences of a) 0.1V and b) 0.2V, at various effective noise temperatures, with bandwidth ($\Delta f$) of 1MHz. At the computer simulations the key length was 700 bits with 500 independent time samples/bit. The asymptote $p = 0.5$ represents perfect security. The evaluation was carried out by the error function, see Equations 19 and 20.

Computer simulations were carried out with $U_{DCA} - U_{DCB} = 0.1V$ and 0.2V. During these tests, $R_L$ and $R_H$ were fixed to 1 and 10 k$\Omega$, respectively. The length of the key was 700 bits. The duration of each BEP was 500 samples (time steps).

The results verify the effectiveness of the attack protocol shown in this paper.

## 6. Defense methods

The attack can be countered using the same defense techniques, as described in [14], namely, cancelling the DC voltages, or by increasing the effective (rms) value of noises by increasing the noise temperature and/or the bandwidth (without exceeding the wave limit [1,2,29]). All methods are the same as those discussed in [14], except for the DC voltages cancellation techniques in which the defense can be conducted in two other ways:

i) Adding variable DC-voltage sources at each side and tuning them to compensate out the parasitic sources. Alternatively tuning them to reach $U_{DCB} - U_{DCA} = 0$.

ii) Naturally, a simplified version would work as well: Adding a single variable DC voltage at one side and tuning it to reach $U_{DCB} - U_{DCA} = 0$ yielding zero DC loop current.

iii) Attaching a capacitor in series to the cable from Alice's end, Bob's end, or from both ends to eliminate DC current in the wire. Note, this maneuver requires great precautions because of its impact on line impedance and the potential information leak.

## 7. Conclusion

This paper generalizes the DC loop current attack introduced in [14]. The generalized scheme makes Eve's work easier. We provided a mathematical analysis and verified the attacks analytically and by computer simulations. We also propose effective defense techniques.

In conclusion, in practical KLJN key exchangers Alice and Bob must carry a DC loop current tests before and during operation and act accordingly (see Section 6).

It is important to note that the general, more expensive defense method of KLJN that is based on in-situ system simulation and comparison with measurements, see Section 4.1 in [16], works efficiently because that alarms for any deviation from the idealized situation, including parasitic DC voltages and currents.